\documentclass[pra,twocolumn,nofootinbib,balancelastpage]{revtex4-1}

\usepackage[hidelinks]{hyperref}

\usepackage{amssymb}
\usepackage{mathtools}
\usepackage{braket}

\usepackage{graphicx}

\usepackage[caption=false]{subfig}

\usepackage{tikz}
\usetikzlibrary{arrows.meta,positioning,fit,decorations,decorations.pathmorphing,fadings}
\usepackage{tikz-3dplot}

\usepackage{pgflibraryarrows}

\usepackage{balance}

\begin{document}

\author{Giuseppe Argentieri}
\affiliation{Department of Physics and Astronomy, University of Exeter, Stocker
  Road, Exeter EX4 4QL, United Kingdom} 
\author{Janet Anders}
\affiliation{Department of Physics and Astronomy, University of Exeter, Stocker
  Road, Exeter EX4 4QL, United Kingdom}

\title{Duality relation for a generalized interferometer}

\begin{abstract}
  It is well known that the Mach-Zender interferometer exhibits a trade-off
  between the a priori which-path knowledge and the visibility of its
  interference pattern. This trade-off is expressed by the inequality
  $\mathcal{P}^2 + \mathcal{V}^2 \leq 1$, constraining the
  \textit{predictability $\mathcal{P}$} and \textit{visibility $\mathcal{V}$} of
  the interferometer. In this paper we extend the Mach-Zender scheme to a setup
  where the central phase shifter is substituted by a generic unitary operator.
  We find that the sum $\mathcal{P}^2 + \mathcal{V}^2$ is in general no longer
  upper bounded by $1$, and that there exists a whole class of interferometers
  such that the full fringe visibility and the full which-way information are
  not mutually exclusive.  We show that $\mathcal{P}^2 + \mathcal{V}^2 \leq L_U$,
  with $1 \leq L_U \leq 2$, and we illustrate how the tight bound $L_U$ depends on
  the choice of the unitary operation $U$ replacing the central phase shifter.

\end{abstract}

\date{\today}

\maketitle

\section{Introduction}

One of the most distinctive aspects of quantum mechanics is the possibility to
adopt two different perspectives which separately highlight the particle-like and
wave-like nature of the same system. Moreover, Bohr's complementarity principle
states that the two descriptions are \textit{mutually exclusive}, in that it is
not possible to perform an experiment which allows the system to exhibit both
its particle and wave properties at the same time~\cite{bohr_quantum_1928}. The
most famous demonstration of this principle is Young's double-slit experiment,
where a beam of particles individually fired through two slits generates a
density distribution on the detection screen which corresponds to the
interference pattern that would be created by a wave. Yet when one tries to
determine through which slit each particle has passed, the interference pattern
is inevitably wiped out. This characteristic is known as ``wave-particle
duality''.

However a quantitative formulation of this principle was first derived only in
1979 by Wootters and Zurek~\cite{wootters_complementarity_1979}: by analysing
Einstein's version of the double-slit experiment, they found that it is possible
to gain partial knowledge of a single photon's path without erasing the
interference pattern. The result was also obtained by
Bartell~\cite{bartell_complementarity_1980} with two alternative variations of
the same apparatus. Greenberger and Yasin extended the analysis to the case of a
two-way neutron interferometer~\cite{greenberger_simultaneous_1988}, and
introduced two relevant quantities that will be adopted in the present work: the
\emph{predictability} $\mathcal{P}$ and the \emph{fringe visibility}
$\mathcal{V}$. The trade-off between the amount of ``path information'' and the
interference sharpness can then be expressed as a constraint on $\mathcal{P}$
and $\mathcal{V}$ in the form of an
inequality:
\begin{equation}\label{eq:engl_ineq}
  \mathcal{P}^2 + \mathcal{V}^2 \leq 1,
\end{equation}
which can be saturated only by some pure states~\cite{englert_fringe_1996}.

Interferometric dualities have been intensively studied in the last
decades~\cite{jaeger_complementarity_1993,jaeger_two_1995,englert_fringe_1996,englert_quantitative_2000,martinez-linares_quality_2004,luis_operational_2004,englert_wave-particle_2008,erez_operational_2009,liu_duality_2009,li_duality_2012,coles_equivalence_2014,jia_influence_2014,bera_duality_2015}:
in these works the predictability, which is based solely on the knowledge of the
preparation of the initial state, is often substituted by another quantity, the
``distinguishability''\cite{englert_fringe_1996}, expressing the degree of
which-way information acquired after the particle has interacted with auxiliary
detectors~\cite{jaeger_complementarity_1993,jaeger_two_1995,englert_fringe_1996,englert_quantitative_2000,martinez-linares_quality_2004,erez_operational_2009}.
These dualities have also been interpreted as constraints over the joint
measurement of pairs of unsharp
observables~\cite{liu_duality_2009,li_duality_2012} and recently it has been
argued that all of them are in fact particular cases of more general entropic
uncertainty relations~\cite{coles_equivalence_2014}.  Experimental validations
of both kinds of dualities have been performed with
neutron~\cite{rauch_static_1984,summhammer_stochastic_1987} and
optical~\cite{liu_relation_2012,tang_revisiting_2013,yan_experimental_2015,heuer_phase-selective_2015}
interferometers, and by various implementations of Wheeler's ``delayed choice''
experiment~\cite{jacques_experimental_2007,jacques_delayed-choice_2008,manning_wheelers_2015}.

It is interesting to notice that while wave-particle duality is usually
considered an inherently quantum feature, there is also evidence of interference
effects in the diffraction of large molecules such as
fullerenes~\cite{arndt_waveparticle_1999}. But interference has been observed at
a much larger scale: a macroscopic droplet bouncing on a vibrating bath can
generate a field of surface waves that couples with it. The field acts as a
pilot wave guiding the droplet as it moves steadily over the
surface~\cite{couder_dynamical_2005}. Couder and Fort have shown that, when
these ``walkers'' pass through a double-slit screen, the distribution of their
scattered trajectories matches the interference pattern of the diffracted pilot
waves~\cite{couder_single-particle_2006}. Furthermore, due the large scale of
the droplets ($\sim1 \mathrm{mm}$) it is possible to observe simultaneously both
their path and the interference of their guiding waves. These striking
similarities between the walkers' behaviour and the quantum wave-particle
duality leave open the question whether the macroscopic (classical) and
microscopic (quantum) worlds are really so different, or if the former can provide
some insight into the latter~\cite{bush_quantum_2010}.

To investigate further the nature of wave-particle duality, here we want to
study an extension of the Mach-Zender interferometric
setup~\cite{englert_fringe_1996}, where the particular unitary operator
representing the middle phase shifter is replaced by an generic unitary $U$. We
find that in this situation the sum $\mathcal{P}^2 + \mathcal{V}^2$ can assume
all values in the range $[0, 2]$. In particular we will show that
\begin{equation}
  \mathcal{P}^2 + \mathcal{V}^2 \leq L_U ,
  \label{eq:duality_2}
\end{equation}
where $L_U$, the maximum value of $\mathcal{P}^2 + \mathcal{V}^2$ taken over all
the possible input states, depends on the choice of the middle unitary $U$, and
$1 \leq L_U \leq 2$. As a consequence the trade-off between the so called
``particle-like'' and ``wave-like'' behaviours disappears.  We will also show that
the transition between the two extremal values is smoothly dependent on the
parameters of the unitary.

This article is structured as follows: in section~\ref{sec:mach-zend-interf} we
briefly describe the Mach-Zender interferometer and the meaning of the two
relevant quantities $\mathcal{P}$ and $\mathcal{V}$. In
section~\ref{sec:gener-interf} we introduce a generalized interferometer by
replacing the middle phase shifter with a generic unitary and in this new scheme
we identify the expressions for predictability and visibility.  In
section~\ref{sec:duality-relation} we show that the generalized setup satisfies
the relation~\eqref{eq:duality_2}, with $1 \leq L_U \leq 2$.  In
section~\ref{sec:vari-ineq-upper} we illustrate the smooth dependence of $L_U$
on the physical parameters of our setup and we interpret the generalized middle
unitary in terms of linear optical transformations. In the last section we
discuss the results and their implications.

\section{The Mach-Zender interferometer}
\label{sec:mach-zend-interf}

\begin{figure}[t!]
  \centering
  \includegraphics[width=1\columnwidth]{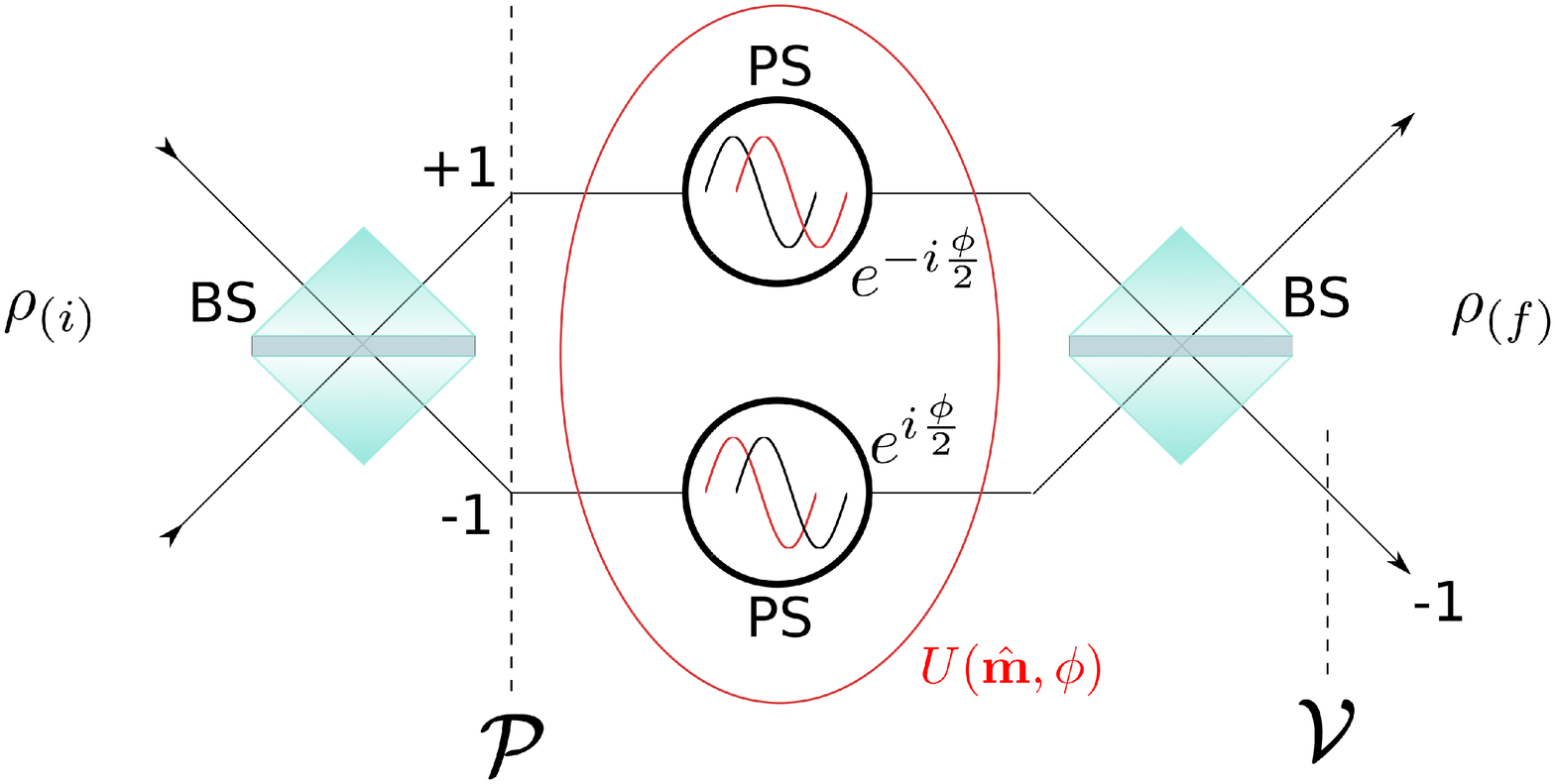}
  \caption{Illustration of the Mach-Zender (M-Z) interferometer as described
    in~\cite{englert_fringe_1996}: after the action of the first beam splitter,
    a differential phase $\phi$  is applied to the two arms by two phase shifters (PS)
    and the beams are then crossed again by another beam splitter. In the
    generalized setup, the phase shifters are replaced by a generic unitary $U$
    which is determined by its rotation axis $\hat{\mathbf{m}}$ and by the angle
    $\phi$ (Eq.~\eqref{eq:unitaries}). The predictability $\mathcal{P}$
    (Eq.~\eqref{eq:Predictability}) is calculated after the first BS, while the
    fringe visibility $\mathcal{V}$ (Eq.~\eqref{eq:visibility_def}) quantifies
    the sharpness of the interference pattern obtained at the end of the
    process when varying $\phi$.}
  \label{fig:MZI}
\end{figure}

In the standard Mach-Zender interferometer (see Fig.~\ref{fig:MZI}), a particle
sent through a 50:50 beam splitter (BS) can take either of two paths. The path
is the relevant degree of freedom, therefore there are two modes, corresponding
to the two paths, and the particle can be effectively described as a $2$-level
system. Identifying the ``which-way'' observable with the Pauli matrix
$\sigma_z$ the Hilbert space is spanned by its two eigenstates:
$\Ket{0}$ and $\Ket{1}$. Their eigenvalues, $+1$ and $-1$, label the two paths. A
convenient representation of the BS is given by the unitary operator
$e^{-i\frac{\pi}{4}\sigma_y}= \frac{1}{\sqrt{2}} \left( \mathbf{1} -i \sigma_y
\right)$,
which maps the path-eigenstates $\Ket{0}$, $\Ket{1}$ onto their equal
superpositions $\Ket{\pm} = (\Ket{0}\pm\Ket{1})/\sqrt{2}$:
\begin{equation*}
  \begin{split}
    e^{-i\frac{\pi}{4}\sigma_y} \Ket{0} = \frac{ \mathbf{1} -i \sigma_y}{\sqrt{2}}
    \Ket{0} = \frac{\Ket{0} + \Ket{1}}{\sqrt{2}} , \\ 
    e^{-i\frac{\pi}{4}\sigma_y} \Ket{1} = \frac{\mathbf{1} -i \sigma_y}{\sqrt{2}}
    \Ket{1} = \frac{\Ket{0} - \Ket{1}}{\sqrt{2}} .
\end{split}
\end{equation*}

Generally the system will be initially prepared in a mixed state, represented by
the density matrix
\begin{equation}\label{eq:rho_i}
  \rho_{(i)} = \frac{\mathbf{1} +
    \mathbf{s}_{(i)}\cdot\hat{\mathbf{\sigma}}}{2}, 
\end{equation}
with Bloch vector $\mathbf{s}_{(i)}$, and where
$\hat{\mathbf{\sigma}}\equiv(\sigma_x, \sigma_y, \sigma_z)$ is the triple of
Pauli's matrices.

The Bloch vector representing the system's state after the action of the first
BS is then
\begin{equation}
  \mathbf{s}_1 = \hat{\mathbf{e}}_y \times \mathbf{s}_{(i)} + s_{(i)y}
  \hat{\mathbf{e}}_y .
\end{equation}
This formula has a straightforward geometric interpretation: $\mathbf{s}_{(i)}$
is rotated by an angle $\pi/2$ about the direction $\hat{\mathbf{e}}_y$ in the
Bloch sphere.

In~\cite{englert_fringe_1996} the which-way predictability was defined as the
\emph{a priori} knowledge of the path taken by the particle after it has passed
through the beam splitter.  The probabilities $w_\pm$ of getting the
values $\pm 1$ upon the measurement of the $\sigma_z$ observable, are given by
\begin{equation}
  w_\pm = \mathrm{Tr} \left[ \frac{\mathbf{1} \pm  \sigma_z}{2} \left(
      e^{-i\frac{\pi}{4}\sigma_y} \frac{\mathbf{1} + \mathbf{s}_{(i)} \cdot
        \hat{\mathbf{\sigma}}}{2} e^{i\frac{\pi}{4}\sigma_y} \right) \right] =\frac{1 \pm s_{1z}}{2}.
\end{equation}

The \textit{predictability} is then defined by
\begin{equation}
  \label{eq:Predictability}
  \mathcal{P}=\lvert w_+-w_-\rvert = \vert s_{1z} \rvert =  \lvert s_{(i)x} \rvert .
\end{equation}

$\mathcal{P}$ estimates the ability to make a correct \textit{guess} of the path
that will be taken. In particular, $\mathcal{P} =1$ corresponds to a full \textit{a priori}
knowledge of the path and it is therefore associated to a particle-like behaviour
(in the classical sense of a system having a definite trajectory). 

After the action of the BS two phase shifters (PS) apply a differential phase
$\phi$ to the two beams, which are then crossed by a second BS, identical to the
first.  Finally, a detector determines whether the particle takes a given exit
path, e.g. the one labeled by $-1$. The interaction with the detector
corresponds to an actual measurement of the observable $\sigma_z$. The procedure
is repeated many times, starting with the same input state $\rho_{(i)}$.  Upon a
large number of measurements of $\sigma_z$ the frequency of the outcome $-1$ is
obtained. This frequency $p$ depends on the applied differential phase
$\phi$. The dependence $p = p(\phi)$ can be reconstructed by performing a set of
many measurement of $\sigma_z$ for different values of $\phi$. As a final result
$p(\phi)$ exhibits a sinusoidal profile~\cite{englert_fringe_1996}. The contrast
of its oscillations is quantified by the fringe visibility $\mathcal{V}$,
defined as the ratio between the amplitude of the oscillations and its average
value or, equivalently, if $I_{\mathrm{max}}$ and $I_{\mathrm{min}}$ are the
maximum and minimum intensity, as
\begin{equation}
  \label{eq:visibility_def}
  \mathcal{V} = \frac{I_{\mathrm{max}}- I_{\mathrm{min}}}{I_{\mathrm{max}}
    +I_{\mathrm{min}}} .
\end{equation}

From their definitions it is clear that the quantities $\mathcal{P}$ and
$\mathcal{V}$ can assume values in the range $[0,1]$.  Nevertheless it turns out
that they are constrained by the inequality~\eqref{eq:engl_ineq}, which includes
all the intermediate situations between the full-defined interferometric pattern
of the pure wave-like case (corresponding to the maximum visibility
$\mathcal{V}=1$) and the full which-way information of the particle-like case
($\mathcal{P}=1$).

In particular it can be shown that
$\mathcal{P}^2 + \mathcal{V}^2 = \lvert \mathbf{s}_{(i)} \vert^2$~(see for
example \cite{englert_fringe_1996}), and therefore the
inequality~\eqref{eq:engl_ineq} is saturated only by pure input states, which
have $\lvert \mathbf{s}_{(i)}\vert ^2 =1$.

\section{Generalized interferometer}
\label{sec:gener-interf}

Inspired by the Mach-Zehnder interferometer here we consider an alternative
scheme where the phase shifter $e^{-i\frac{\phi}{2}\sigma_z}$ is replaced by a
general unitary transformation (see Fig.~\ref{fig:MZI})
\begin{equation}\label{eq:unitaries}
  U = e^{-i\hat{\mathbf{m}}\cdot\hat{\mathbf{\sigma}}\,\phi/2} .
\end{equation}
Here the rotation axis is determined by the unit vector $\hat{\mathbf{m}}$ and
the rotation angle given by $\phi$. We also implement the second BS with the
inverse of the first one, that is $e^{i\sigma_y\pi/4}$. With this choice, in
absence of any middle transformation, $\sigma_z$'s eigenstates are mapped back
into themselves.

The system's state now undergoes the overall unitary
$V \equiv e^{i\frac{\pi}{4}\sigma_y} U e^{-i\frac{\pi}{4}\sigma_y}$, which can
be written as a function of the parameters $\phi$ and $\hat{\mathbf{m}}$:
\begin{equation}\label{eq:V}
  V = e^{-i  \frac{\phi}{2} (-m_z \sigma_x + m_y \sigma_y + m_x \sigma_z ) } =
  e^{-i \mathbf{t} \cdot \hat{\sigma} \frac{\phi}{2}} ,
\end{equation}
where the $3$-dimensional unit vector $\mathbf{t}(\hat{\mathbf{m}})$ takes
the form
\begin{equation}
  \mathbf{t}(\hat{\mathbf{m}}) = -m_z \hat{\mathbf{e}}_x + m_y
  \hat{\mathbf{e}}_y + m_x\hat{\mathbf{e}}_z . 
\end{equation}
The final state 
\begin{equation}\label{eq:rho_fin}
  \rho_{(f)} = V \rho_{(i)} V^\dagger = \frac{ \mathbf{1}+\mathbf{s}_{(f)}\cdot\hat{\mathbf{\sigma}} }{2},
\end{equation}
is then determined by the Bloch vector
\begin{equation}\label{eq:s_f}
  \mathbf{s}_{(f)} = \cos\phi\, \mathbf{s}_{(i)} + \sin\phi\,
  (\mathbf{t}\times\mathbf{s}_{(i)}) + (  1-\cos\phi)\,
  (\mathbf{t}\cdot\mathbf{s}_{(i)}) \mathbf{t} . 
\end{equation}

After the qubit has passed through the second BS, the path measurement
$\sigma_z$ is finally performed. The probability of getting one of the two
outcomes, say $-1$, is the expectation value of the projector
$(\mathbf{1}-\sigma_z)/2$ evaluated on the final state $\mathbf{\rho}_{(f)}$:
\begin{equation}
  p = \mathrm{Tr} \left[
    \left(\frac{\mathbf{1}-\sigma_z}{2}\right)
    \left(\frac{\mathbf{1}+\mathbf{s}_{(f)}\cdot\hat{\mathbf{\sigma}})}{2}
    \right) \right] .
\end{equation}

By using~(\ref{eq:s_f}), this probability can be expressed as a function of the
phase $\phi$ applied by the \mbox{unitary $U$:}
\begin{equation}
  \begin{split} 
    p = p(\phi) = \frac{1}{2} &\Big\{ 1 - (\mathbf{t}\cdot\mathbf{s}_{(i)}) t_z
    - \left[ s_{(i)z} - (\mathbf{t}\cdot\mathbf{s}_{(i)}) t_z \right] \cos\phi
    \\ 
    & - \left[ (\mathbf{t}\times\mathbf{s}_{(i)})_z \right] \sin\phi \Big\} .
  \end{split}
  \label{eq:p_phi} 
\end{equation}

The \textit{visibility} of a sinusoidal interferometric pattern is defined as
in~\eqref{eq:visibility_def}, which leads to the following result\footnote{The
  denominator vanishes only when $(\mathbf{t}\cdot\mathbf{s}_{(i)}) t_z = 1$,
  i.e. only when $\mathbf{s}_{(i)} = \hat{\mathbf{e}}_z = \pm \mathbf{t}$, but
  in this case also the probability $p(\phi)$~\eqref{eq:p_phi} vanishes and the
  visibility is $0$.} :
\begin{equation}
  \label{eq:visibility}
  \mathcal{V} = \sqrt{\frac{\left[ s_{(i)z} - (\mathbf{t}\cdot\mathbf{s}_{(i)})  t_z \right]^2 + \left[
        (\mathbf{t}\times\mathbf{s}_{(i)})_z  \right]^2 }{ \left[1 -
      (\mathbf{t}\cdot\mathbf{s}_{(i)}) t_z \right]^2}} .   
\end{equation}

So far the quantities $\mathcal{P}$ and $\mathcal{V}$ have been derived in the
most general case, where no assumption has been made on the vector
$\hat{\mathbf{m}}$ appearing in~\eqref{eq:unitaries}. In the following section
we show that the sum $\mathcal{P}^2 + \mathcal{V}^2$ is crucially dependent on
the mutual orientation of $\hat{\mathbf{e}}_y$ and $\hat{\mathbf{m}}$ (i.e. of
the vectors defining the BS unitary and the middle unitary, respectively) as
well as on the angle between $\hat{\mathbf{m}}$ and $\hat{\mathbf{e}}_z$, which
is the axis identifying the ``which-path'' observable. In particular, the
inequality~\eqref{eq:engl_ineq} is no longer valid in the generalized
interferometer.

\section{Duality relation}
\label{sec:duality-relation}

First, we briefly discuss a trivial case for the system's input state, i.e. the
\textit{totally depolarized} state $\rho_{(i)} = \mathbf{1}/2$, corresponding to
$\mathbf{s}_{(i)} = 0$. If $\mathbf{s}_{(i)}= 0$ the system's state remains
unaltered under any unitary, therefore $\mathbf{s}_1 = 0$ and
$\mathbf{s}_{(f)} = 0$ imply that both the
predictability~\eqref{eq:Predictability} and the probability
$p(\phi)$~\eqref{eq:p_phi} vanish. As one would expect for the maximally
entropic state, one obtains $\mathcal{P}^2 + \mathcal{V}^2 = 0$.

Another interesting case is the pure input state
\mbox{$\mathbf{s}_{(i)} =\hat{\mathbf{e}}_z$}. This is mapped by the BS onto the
Bloch vector $\hat{\mathbf{e}}_x$, whose density matrix represents the projector
onto a 50:50 superposition of the two which-way eigenstates, that is
$(\Ket{0} + \Ket{1})/\sqrt{2}$. Therefore, in agreement with
Eq.~\eqref{eq:Predictability}, $\mathcal{P} =0$. It is easy to check that the
visibility instead is $\mathcal{V}=1$.

Here we want to show that our generalized setup implies that, in a large number
of cases, the sum $\mathcal{P}^2 + \mathcal{V}^2$ can be higher than $1$ and, in
fact, it can easily reach its maximum value $2$. In order to simplify our
analysis we pick now as initial states the Bloch vectors
$\mathbf{s}_{(i)} = s_x \hat{\mathbf{e}}_x$, with \mbox{$0 \leq s_x \leq 1$}. As
they are mapped onto the $z$-axis, the predictability can assume any value from
$0$ to $1$.

\begin{figure}[]
  \tdplotsetmaincoords{55}{100}
  \begin{tikzpicture}[tdplot_main_coords,fill=gray]
    
    \def\R{2.5} 
    \coordinate (0) at (0,0,0); 
    
    \filldraw[ball color=white,tdplot_screen_coords,fill opacity=0.8] (0,0)
    circle (\R); \tdplotdrawarc[thin]{(0)}{\R}{-80}{100}{}{}; 
    \tdplotdrawarc[thin,dashed]{(0)}{\R}{100}{280}{}{};

    \draw[thick,color=blue,-Stealth] (0) -- (\R,0,0) node[anchor=north]
    {$\hat{\mathbf{e}}_y$};

    \tdplotsetcoord{s}{.66*\R}{90}{75}; \draw[thick,color=orange,-Stealth]
    (0) -- (s) node[anchor=north] {$\mathbf{s}_{(i)}$};

    \tdplotsetcoord{ss}{\R}{90}{75}; \draw[dashed,color=orange] (s) -- (ss);

    \draw[thick,-Stealth] (0) -- (0,0,\R) node[anchor=east]
    {$\hat{\mathbf{e}}_z$}; \draw[thick] (0,0,-\R) {} ;

    \draw[thick,-Stealth] (0) -- (0,-\R,0) node[anchor=north west]
    {$\hat{\mathbf{e}}_x$};

    \tdplotsetcoord{m}{\R}{50}{45} 
    \draw[thick,color=red,-Stealth] (0) -- (m) node[anchor=west]
    {$\hat{\mathbf{m}}$}; \draw[thick,dashed,color=red,-Stealth] (0) -- (mz)
    node[anchor=south east] {$\mathbf{m}_\perp$};
    \draw[thick,dashed,color=red,-Stealth] (0) -- (mxy) node[anchor=north]
    {$\mathbf{m}_P$}; \draw[dashed,thin,color=gray] (mz) -- (m);
    \draw[dashed,thin,color=gray] (mxy) -- (m);
    
    \tdplotdrawarc[thin, color=gray,
    -Stealth]{(0)}{.4*\R}{-90}{45}{anchor=north}{$\xi$}; 
    
    \tdplotsetthetaplanecoords{45} \tdplotdrawarc[tdplot_rotated_coords,
    thin, color=gray,
    -Stealth]{(0)}{.5*\R}{0}{50}{anchor=north}{$\Theta$}; 
  \end{tikzpicture}
  \caption{The decomposition of the unit vector $\hat{\mathbf{m}}$ into its
    projection $\mathbf{m}_P$ onto the plane $(x,y)$ and its orthogonal
    component $\mathbf{m}_\perp$. The \textit{visibility} can be expressed as a
    function of the polar variables $\Theta$ and $\xi$.}
  \label{fig:Bloch_sphere_1}  
\end{figure}
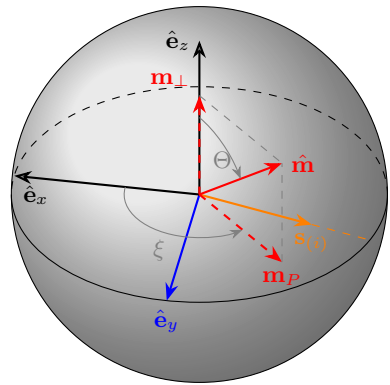

The unit vector $\hat{\mathbf{m}}$ can be decomposed into its projected
component $\mathbf{m}_P$ on the $(x,y)$ plane and the orthogonal component
$\mathbf{m}_\perp$ (see Fig.~\ref{fig:Bloch_sphere_1}); then it can be written
in spherical coordinates:
\begin{equation}\label{eq:m_dec}
  \hat{\mathbf{m}} = \mathbf{m}_{\perp} + \mathbf{m}_{P} =   \cos\Theta \,
  \hat{\mathbf{e}}_z + \sin\Theta \left( \cos\xi \,  \hat{\mathbf{e}}_x + \sin\xi \,
    \hat{\mathbf{e}}_y \right) .  
\end{equation}

In this way we can express $\mathcal{P}^2 + \mathcal{V}^2$ in spherical coordinates
through~\eqref{eq:Predictability}, \eqref{eq:visibility}
and~\eqref{eq:m_dec}. We call this function $F$:
\begin{equation}
  \label{eq:F_tilde}
  F(s_x, \Theta, \xi ) = \lvert s_x \rvert^2 \left( 1+ \frac{ \sin^2\Theta ( \cos^2\Theta \cos^2\xi
      + \sin^2\xi )}{ (1 + s_x \sin\Theta \cos\Theta \cos\xi)^{2}} \right) ,
\end{equation}
and notice that it spans the whole range of values \mbox{$0 \leq F \leq 2$}. 

\begin{figure}[]
  \includegraphics[width=.45\textwidth]{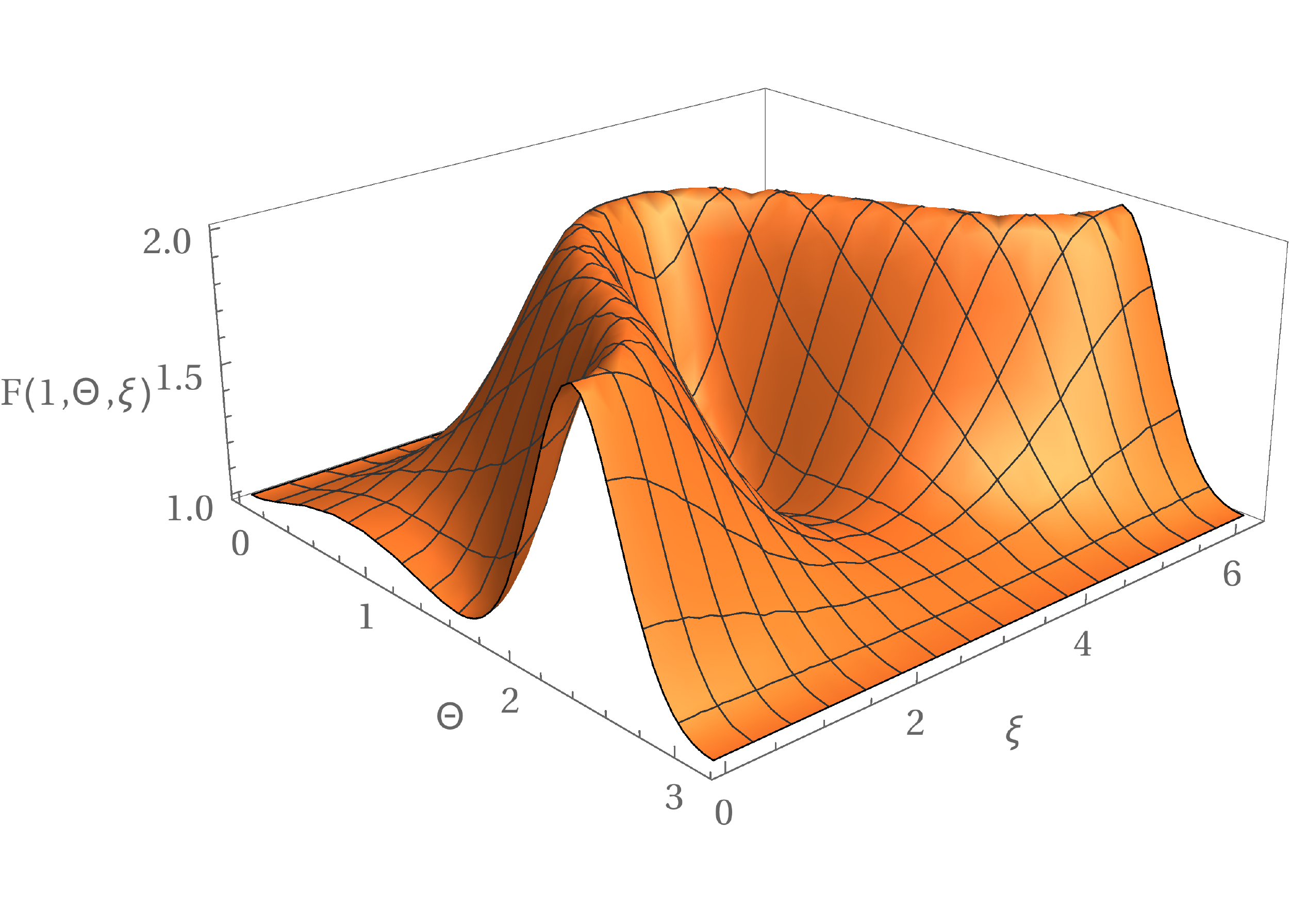}
  \caption{Plot of the function $F(1, \Theta, \xi) = \mathcal{P}^2+\mathcal{V}^2$ corresponding to
    the choice $\mathbf{s}_{(i)}= \hat{\mathbf{e}}_x$. Here $\Theta$ is the
    angle between the vectors $\hat{\mathbf{m}}$ and $\hat{\mathbf{e}}_z$, and
    $\xi$ the angle between $\hat{\mathbf{e}}_x$ and the projection
    $\mathbf{m}_P$ of $\hat{\mathbf{m}}$ on the plane $(x,y)$. There are
    configurations of the vector $\hat{\mathbf{m}}$ such that the quantity
    $F = \mathcal{P}^2 + \mathcal{V}^2$ can be larger than $1$.}
  \label{fig:F_Theta}
\end{figure}

If we take as input the pure state $\mathbf{s}_{(i)}=\hat{\mathbf{e}}_x$ then
the sum will assume the simpler form:
\begin{equation} \label{eq:F_pol}
  F(1, \Theta, \xi ) = \left( 1 + \frac{\sin^2\Theta \cos^2\Theta \cos^2\xi +
      \sin^2\Theta \sin^2\xi}{(1 + \sin\Theta \cos\Theta \cos\xi)^{2}}
  \right) .
\end{equation}
In Fig.~\ref{fig:F_Theta} this function is plotted. It is evident that the
inequality
\begin{equation}
  \label{eq:ineq}
  \mathcal{P}^2 + \mathcal{V}^2 \leq 2 
\end{equation}
is saturated for a variety of different orientations of the vector
$\hat{\mathbf{m}}$. Moreover, both the inequalities~\eqref{eq:engl_ineq}
and~\eqref{eq:ineq} can only be saturated by the two input states
\mbox{$\mathbf{s}_{(i)} = \pm \hat{\mathbf{e}}_x$}, since the predictability is
given by
\mbox{$\mathcal{P}=\lvert \mathbf{s}_{(i)} \cdot \hat{\mathbf{e}}_x \rvert$}.

It is interesting to observe what condition on $\hat{\mathbf{m}}$ determines an
upper bound equal to $1$: the second term in~\eqref{eq:F_tilde} vanishes if
either $\Theta \in \{0,\pi\}$, which means that
$\hat{\mathbf{m}}=\pm \hat{\mathbf{e}}_z$ and therefore the unitary
$U = e^{-i\phi\sigma_z/2}$ is a phase shifter as in~\cite{englert_fringe_1996},
or if $\xi=0$ and $\cos\Theta = 0$, which means
$\hat{\mathbf{m}}=\pm \hat{\mathbf{e}}_x$, that is if the central unitary is
another beam splitter\footnote{This is easily checked by applying
  $e^{-i\phi\sigma_x/2}$ to $\sigma_z$'s eigenstates $\Ket{0}$, $\Ket{1}$.}:
$U=e^{-i\phi\sigma_x/2}$.

However, orthogonality between the two unit vectors $\hat{\mathbf{m}}$ and
$\hat{\mathbf{e}}_y$ is not sufficient to ensure that
$\mathcal{P}^2 + \mathcal{V}^2 \leq 1$ in this scenario, since one can take
$\hat{\mathbf{m}}$ in the plane $(x,z)$ such that $F \geq 1$: e.g.  if
$\hat{\mathbf{m}} = (\hat{\mathbf{e}}_x+\hat{\mathbf{e}}_z)/\sqrt{2}$, then
$\Theta=\pi/4$, $\xi=0$ imply
$F(s_x,\Theta,\xi) = \lvert s_x \rvert^2 (1 + 1/(2+s_x)^2)$ which is larger than
$1$ for the pure state $\mathbf{s}_{(i)} = \hat{\mathbf{e}}_x$.  The reason is
that, unlike the Mach-Zender interferometer analysed
in~\cite{englert_fringe_1996}, the observable which is being measured,
$\sigma_z$, is now different from the operator
$\hat{\mathbf{m}}\cdot\hat{\mathbf{\sigma}}$ defining the unitary $U$.

The duality relation~\eqref{eq:ineq} has an important consequence: if one
considers generalized interferometric setups, where the middle unitary is not a
phase-shifter, then it is possible to achieve a \textit{perfect} which-way
knowledge ($\mathcal{P}=1$) without destroying the interference pattern
($\mathcal{V} \neq 0$). And viceversa: a \textit{full} fringe visibiliy
($\mathcal{V}=1$) is no longer incompatible with a high degree of
predictability.

\section{Variation of the inequality's upper bound}
\label{sec:vari-ineq-upper}

In order to illustrate the dependence of the visibility~\eqref{eq:visibility} on
the mutual orientation of $\hat{\mathbf{e}}_y$ and $\hat{\mathbf{m}}$, we consider again
the pure initial state $\mathbf{s}_{(i)} = \hat{\mathbf{e}}_x$ and recast the
sum $\mathcal{P}^2+\mathcal{V}^2$ in terms of the $(x,y,z)$ coordinates of the
vector $\hat{\mathbf{m}}$.  From Eq.~\eqref{eq:m_dec}
\begin{equation}
  m_x =   \sin\Theta \cos\xi, \quad m_y = \sin\Theta \sin\xi,    \quad  m_z = \cos\Theta .
\end{equation}

Therefore Eq.~\eqref{eq:F_pol} can be rewritten, reminding that
$\hat{\mathbf{m}}$ is a unit vector ($m_x^2 + m_y^2 + m_z^2 =1$), as a function
of $m_x$ and $m_z$:
\begin{equation}\label{eq:f_mx_mz}
  \mathcal{P}^2+\mathcal{V}^2 = f(m_x, m_z) \equiv  \left[ 1 + \frac{(1 -
      m_x^2)(1 - m_z^2)}{(1+m_zm_x)^2}  \right] .
\end{equation}

\begin{figure}[]
  \includegraphics[width=1\columnwidth
  ]{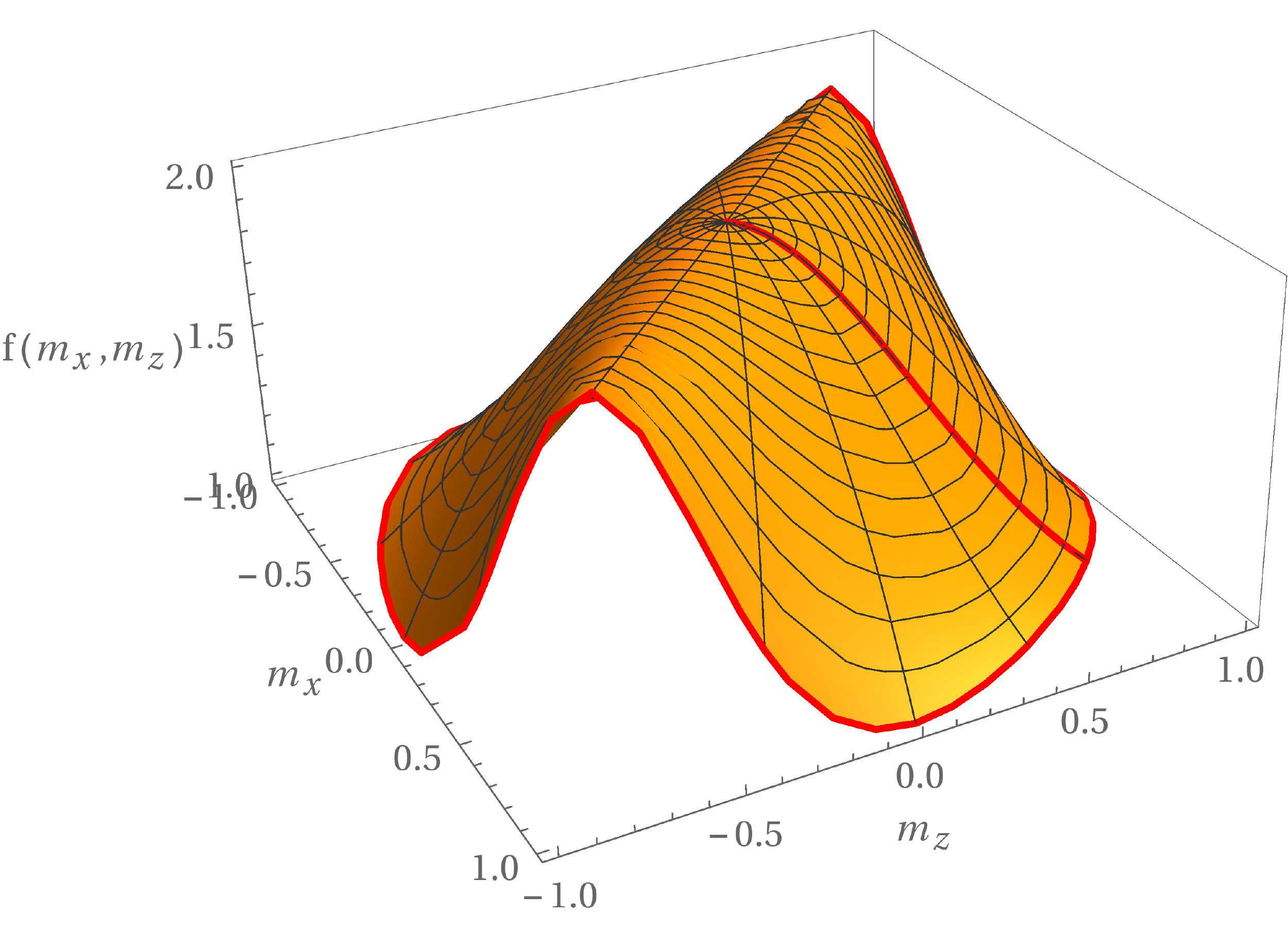}
  \caption{Plot of the function $f(m_x,m_z) $~\eqref{eq:f_mx_mz} for
    \mbox{$\mathbf{s}_{(i)}=\hat{\mathbf{e}}_x$}. The red radial path drawn over the
    manifold is the profile of the one-dimensional function \mbox{$m_x \mapsto f(m_x,m_x)$. }}
  \label{fig:f_mx_mz}
\end{figure}

It can be shown that this function has no local \emph{isolated} extremal points
but it assumes its maximum value $1$ on the line $m_z=-m_x$ (see
Fig.~\ref{fig:f_mx_mz}). On the other hand, if we restrict the study of
$f(m_x,m_z)$ to the subdomain
\mbox{$\mathcal{L} =\{ (m_x, m_x): m_x \in [0,\frac{1}{\sqrt{2}}] \}$}, the
profile of the one-dimensional function
$[0,\frac{1}{\sqrt{2}}] \ni m_x \mapsto f(m_x, m_x)$ is monotonic with maximum
and mininum values $2$ and $\frac{10}{9}$. In Fig.~\ref{fig:f_mx_mz} this
profile is shown in red.

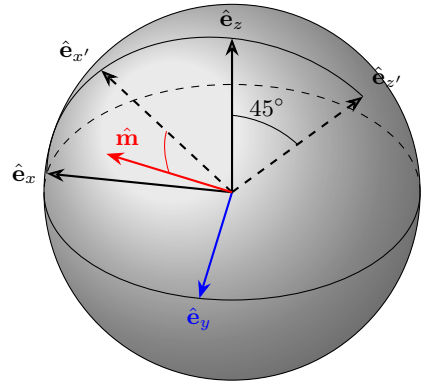
\begin{figure}[!h]
  \tdplotsetmaincoords{55}{100}
  \vspace{10pt}
  \begin{tikzpicture}[tdplot_main_coords,fill=gray]
    \def\R{2.5} 
    \def\L{.5773502692*\R} 
    \def\LL{.7071067812*\R} 

    \coordinate (0) at (0,0,0); 
    
    \filldraw[ball color=white,tdplot_screen_coords,fill opacity=0.8] (0,0)
    circle (\R); \tdplotdrawarc[thin]{(0)}{\R}{-80}{100}{}{}; 
    \tdplotdrawarc[thin,dashed]{(0)}{\R}{100}{280}{}{};

    \draw[thick,color=blue,-Stealth] (0) -- (\R,0,0) node[anchor=north]
    {$\hat{\mathbf{e}}_y$};

    \draw[thick,-Stealth] (0) -- (0,0,\R) node[anchor=south]
    {$\hat{\mathbf{e}}_z$};

    \draw[thick,-Stealth] (0) -- (0,-\R,0) node[anchor=east]
    {$\hat{\mathbf{e}}_x$};
    
    \tdplotsetrotatedcoords{-90}{-45}{-90}

    \draw[thick,dashed,-Stealth,tdplot_rotated_coords] (0) -- (0, 0, \R)
    node[anchor=south west] {$\hat{\mathbf{e}}_{z'}$};
    \draw[thick,dashed,-Stealth,tdplot_rotated_coords] (0) -- (0, \R, 0)
    node[anchor=south east] {$\hat{\mathbf{e}}_{x'}$};

    \tdplotdefinepoints(0,0,0)(0,0,1)(0,\LL,\LL)
    \tdplotdrawpolytopearc[thin]{.5*\R}{anchor=south}{$45^\circ$}

    \draw[thick,red,-Stealth] (0) -- (\L,-\L,\L) node[anchor=south west]
    {$\hat{\mathbf{m}}$};

    \tdplotdefinepoints(0,0,0)(0,-\LL,\LL)(\L,-\L,\L)
    \tdplotdrawpolytopearc[thin,color=red]{.5*\R}{}{}
    
    \tdplotdefinepoints(0,0,0)(0,-\R,0)(0,\LL,\LL)
    \tdplotdrawpolytopearc[thin]{\R}{}{}
  \end{tikzpicture}
  \caption{The vector $\hat{\mathbf{m}}$ is restricted to the
    $(\hat{\mathbf{e}}_y,\hat{\mathbf{e}}_{x'})$ plane. This constraint
    simplifies the discussion and allows an immediate display of the transition
    from a situation, $\hat{\mathbf{m}} \perp \hat{\mathbf{e}}_y$, where the
    quantity $\mathcal{P}^2+\mathcal{V}^2 $ assumes its minimum (but it is still
    greater than 1) to a situation,
    $\hat{\mathbf{m}} \parallel \hat{\mathbf{e}}_y$, where the the full
    knowledge of both observables at the same time is allowed
    ($\mathcal{P}^2+\mathcal{V}^2 \le 2$) .}
  \label{fig:Bloch_sphere_2}
\end{figure}

A suitable change of coordinates provides a more vivid picture of the mentioned
transition (see Fig.~\ref{fig:Bloch_sphere_2}): upon a $-45^\circ$ rotation
about the $y$ axis the orthonormal basis
$(\hat{\mathbf{e}}_x, \hat{\mathbf{e}}_y, \hat{\mathbf{e}}_z)$ is mapped onto
\begin{equation}\label{eq:rot_bases}
  \hat{\mathbf{e}}_{x'} =
  \frac{\hat{\mathbf{e}}_x+\hat{\mathbf{e}}_z}{\sqrt{2}} , \quad
  \hat{\mathbf{e}}_{y'} = \hat{\mathbf{e}}_y, \quad \hat{\mathbf{e}}_{z'} =
  \frac{-\hat{\mathbf{e}}_x + \hat{\mathbf{e}}_z }{\sqrt{2}} .
\end{equation}

In this coordinates system the $1$-dimensional manifold $\mathcal{L}$ is
parametrized as $\mathcal{L}= \{ (m_{x'},0): m_{x'} \in [0,1] \}$. As the
$m_{x'}$ component varies from $0$ to $1$, the unit vector $\hat{\mathbf{m}}$
rotates from $\hat{\mathbf{e}}_y$ to $\hat{\mathbf{e}}_{x'}$:
$\hat{\mathbf{m}}\equiv\hat{\mathbf{e}}_y$ means that the overall unitary
operator~\eqref{eq:V} is actually equivalent to $U$ and the visibility is at
its maximum, $1$, while when the two vectors $\hat{\mathbf{m}}$ and
$\hat{\mathbf{e}}_y$ are orthogonal ($m_{x'}=1$) then the fringe pattern assumes
its minimum.

\begin{figure}
   \centering
   \includegraphics[width=1\columnwidth]{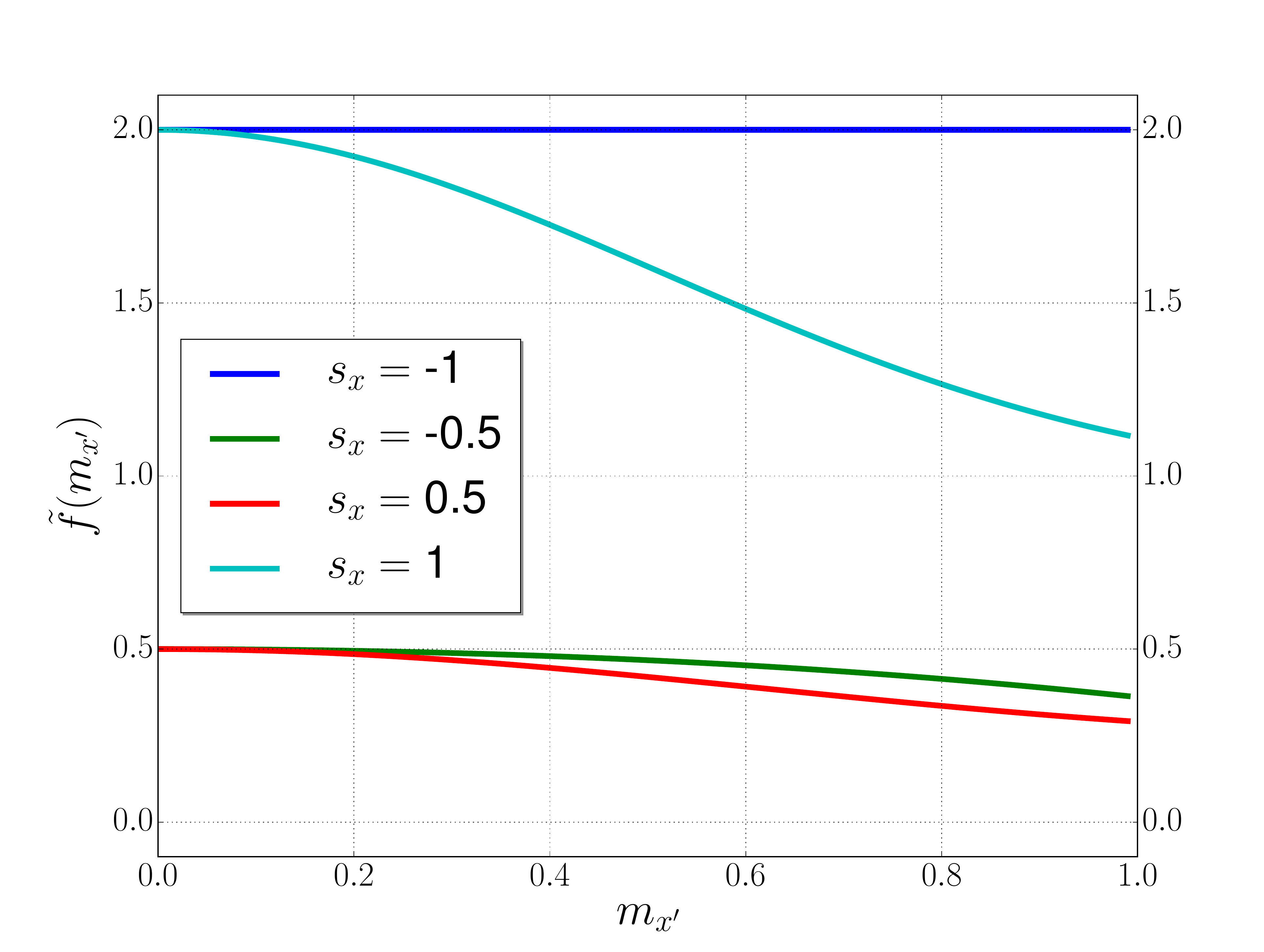}
   \caption{The function $\tilde{f}(s_x, m_{x'})$, Eq.~\eqref{eq:f_tilde}, plotted for
     different initial states $\mathbf{s}_{(i)} = s_x\hat{\mathbf{e}}_x$, shows
     how the interferometric duality's upper bound is $2$, corresponding to
     $\mathcal{P} = \mathcal{V} = 1$. The plot with $s_x=1$ corresponds to the
     red radial profile shown in Fig.~\ref{fig:f_mx_mz}.}
   \label{fig:f_mx}
 \end{figure}

We can also reintroduce the dependence on the initial state
$\mathbf{s}_{(i)} = s_x \hat{\mathbf{e}}_x$ as in~\eqref{eq:F_tilde}. In the new
parameters $F$ will become a function \mbox{$\tilde{f}(s_x, m_{x'})$} of $s_x$
and $m_x'$:
\begin{multline}\label{eq:f_tilde}
  \tilde{f}(s_x, m_{x'}) = \mathcal{P}^2(s_x) +\mathcal{V}^2(s_x, m_{x'}) =\\
  s_x^2 + s_x^2 \frac{\left(1 -
      (\frac{m_{x'}}{\sqrt{2}})^2\right)^2}{\left(1+s_x
      (\frac{m_{x'}}{\sqrt{2}})^2\right)^2} . 
\end{multline}
$\tilde{f}(s_x,m_{x'})$is plotted in Fig.~\ref{fig:f_mx} for different values of
$s_x$. 

So far we have considered 50:50 BS represented by the unitary
$e^{-i\pi\sigma_y/4}$. One can generalize it to an operator of the form
$e^{-i\omega\sigma_y/2}$.  The angle $\omega$ can now vary, and any value
other than $\pi/2$ corresponds to an unbalanced BS. In fact, if
$t_\pm$ and $r_\pm$ are the transmissivity and reflectivity coefficients, then
the unitary
\begin{equation}
  e^{-i\frac{\omega}{2}\sigma_y} = 
  \begin{pmatrix}
    t_+(\omega) & r_-(\omega) \\
    r_+(\omega) & t_-(\omega) 
  \end{pmatrix} 
  =
  \begin{pmatrix}
    \cos\frac{\omega}{2} & -\sin\frac{\omega}{2}  \\
    \sin\frac{\omega}{2} & \cos\frac{\omega}{2}
  \end{pmatrix} ,
\end{equation}
maps the initial state represented by the Bloch vector $\hat{\mathbf{s}}_{(i)}$
onto a state represented by
\begin{equation}
  \hat{\mathbf{s}}_1 = \cos\omega \hat{\mathbf{s}}_{(i)} + \sin\omega
  (\hat{\mathbf{e}}_y \times \hat{\mathbf{s}}_{(i)}) + (1-\cos\omega) s_{(i)y}
  \hat{\mathbf{e}}_y .
\end{equation}

Consequently the probabilities for the particle to take either path after
passing through the BS are given by
\begin{equation}
  w_{\pm} =   \frac{1 \pm s_{(1)z}}{2} = \frac{1}{2} ( 1 \pm s_{(i)z} \cos\omega \mp
  s_{(i)x}\sin\omega ).
\end{equation}

However, varying $\omega$ does not change the main result: if we still consider
a preparation state $\mathbf{s}_{(i)} = s_x \hat{\mathbf{e}}_x$, we get the
following expression for the sum $\mathcal{P}^2+\mathcal{V}^2$
\begin{widetext}
  \begin{equation}
    \mathcal{P}^2 + \mathcal{V}^2 = s_x^2 \left( \sin^2 \omega + \frac{\frac{
          1}{4} \left( \sin 2\Theta \cos 2\omega \sin\xi + \sin 2\omega \cos 2\Theta +
          \sin 2\omega \sin^2 \Theta \cos^2 \xi  \right)^2 + \sin^2 \Theta \cos^2 \xi }{
        \left[ 1 - \frac{s_x}{2} \left( \sin 2\Theta \cos 2\omega \sin\xi + \sin 2\omega
            \cos 2\Theta + \sin 2\omega \sin^2 \Theta \cos^2 \xi  \right) \right]^2  } \right) ,
  \end{equation}
\end{widetext}
which still spans all the range of values
$0 \leq \mathcal{P}^2 + \mathcal{V}^2 \leq 2$ and is saturated by the pure state
$s_x=1$. Similarly, one can relax the condition
$\mathbf{s}_{(i)} = s_x \hat{\mathbf{e}}_x$ to an arbitrary Bloch vector without
any substantial difference.

The most immediate physical visualization of the generalized interferometer is a
qubit undergoing a sequence of three precessions about the two axes
$\hat{\mathbf{e}}_y$ and $\hat{\mathbf{m}}$, where the third rotation is the
inverse of the first. It can be realized by applying sequentially three magnetic
fields to a $1/2$-spin with a non-null magnetic moment. If one chooses the
observable of interest to be the projection along the direction
$\hat{\mathbf{e}}_z$, then the setup can be thought of as a binary
interferometer: the two eigenvalues $\pm 1/2$ correspond to the two ``paths''.

 In our generalized scheme the unitary $U$ is no longer a phase-shifter. Indeed,
 a generic $2 \times 2$ unitary matrix can be factorized as
\begin{equation}\label{eq:matrix_factorization}
  U= e^{i\varphi} 
  \begin{pmatrix}
    e^{i \psi} & 0 \\
    0 & e^{-i \psi} 
  \end{pmatrix}
  \begin{pmatrix}
    \cos\chi & \sin\chi \\
    -\sin\chi & \cos\chi
  \end{pmatrix}
  \begin{pmatrix}
    e^{i\Delta} &0 \\
    0 & e^{-i\Delta} \\
  \end{pmatrix},
\end{equation}
which means that, in our coordinates representation, it can be thought of as a
composition of two phase shifts and a beam splitter (up to an overall phase
factor $e^{i\varphi}$) according to the sequence
\begin{equation}\label{eq:unitary_factorization}
  U = e^{i\psi\sigma_z} e^{i\chi\sigma_y} e^{i\Delta \sigma_z} .
\end{equation}

Therefore the action of the optical operation represented by $U_2$ consists of a
first differential phase shift of an amount $2\Delta$, followed by a beam
splitter and then by another $2\psi$-phase shift. Nevertheless our optical setup
keeps the structure of an interferometer since, as we have already pointed out,
the initial and final unitaries still act as two beam splitters. 

\vspace{10pt}

\section{Conclusions}
\label{sec:conclus}

In this work we have analysed a linear optical setup which can be considered as
a natural generalization of the standard Mach-Zender interferometer. This setup
is formally implemented by three consecutive unitary operators: the first and
the last one still represent two beam splitters, while the middle one is a
generic unitary.

We have derived a new duality relation involving two quantities, the
\textit{fringe visibility} $\mathcal{V}$ and the \textit{predictability}
$\mathcal{P}$, which are widely used to characterize the sharpness of the
interference pattern and the amount of available information about the
which-path observable, respectively. This duality relation is an inequality of
the form
\begin{equation*}
  \mathcal{P}^2 + \mathcal{V}^2 \leq L_U  ,
\end{equation*}
but its upper bound
$L_U = \max_{\rho_{(i)}} \{ \mathcal{P}^2 + \mathcal{V}^2\}$, unlike the
Mach-Zender interferometer, can assume values between $1$ and $2$. We have also
shown how this upper bound depends on the parameters of the generalized
interferometer.

Some considerations are in order. On the one hand the
inequality~\eqref{eq:engl_ineq} is commonly interpreted as the expression of a
genuinely quantum property, the wave-particle duality. Nevertheless it is known
that classical macroscopic objects can show a very similar
behaviour~\cite{bush_quantum_2010}. On the other hand our findings indicate that
the trade-off between the so called ``particle-like'' and ``wave-like'' behaviour
is in fact the property of a specific kind of interferometer and that it
disappears when considering more general schemes of two-way interferometers.  We
think that the framework introduced in section~\ref{sec:gener-interf}, where the
parameters of the central unitary are not fixed, can be adapted to different
kinds of situations and may inspire the study of new types of interferometric
setups and experiments. In particular, Eq.~\eqref{eq:unitary_factorization}
indicates that the abstract unitary $U$ can be operationally implemented as a
sequence of optical transformations made of beam-splitters and phase-shifters.

Finally we believe that our work poses an interesting issue which is worth to be
further investigated both theoretically and experimentally.  If we accept that
wave-particle duality is a distinctive feature of quantum mechanics (according
Feynman the only ``mystery'' of quantum mechanics) then our inequality seems to
suggest that, in the best case, duality does not imply incompatibility as in the
common interpretation of Eq.~\eqref{eq:engl_ineq}. Another possible consequence
is that all these relations are not actually able to capture the ``mystery'',
namely that quantities like $\mathcal{P}$ and $\mathcal{V}$ are not valid
witnesses of the dual nature of quantum systems.

\bibliography{duality}

\begin{thebibliography}{30}%
\makeatletter
\providecommand \@ifxundefined [1]{%
 \@ifx{#1\undefined}
}%
\providecommand \@ifnum [1]{%
 \ifnum #1\expandafter \@firstoftwo
 \else \expandafter \@secondoftwo
 \fi
}%
\providecommand \@ifx [1]{%
 \ifx #1\expandafter \@firstoftwo
 \else \expandafter \@secondoftwo
 \fi
}%
\providecommand \natexlab [1]{#1}%
\providecommand \enquote  [1]{``#1''}%
\providecommand \bibnamefont  [1]{#1}%
\providecommand \bibfnamefont [1]{#1}%
\providecommand \citenamefont [1]{#1}%
\providecommand \href@noop [0]{\@secondoftwo}%
\providecommand \href [0]{\begingroup \@sanitize@url \@href}%
\providecommand \@href[1]{\@@startlink{#1}\@@href}%
\providecommand \@@href[1]{\endgroup#1\@@endlink}%
\providecommand \@sanitize@url [0]{\catcode `\\12\catcode `\$12\catcode
  `\&12\catcode `\#12\catcode `\^12\catcode `\_12\catcode `\%12\relax}%
\providecommand \@@startlink[1]{}%
\providecommand \@@endlink[0]{}%
\providecommand \url  [0]{\begingroup\@sanitize@url \@url }%
\providecommand \@url [1]{\endgroup\@href {#1}{\urlprefix }}%
\providecommand \urlprefix  [0]{URL }%
\providecommand \Eprint [0]{\href }%
\providecommand \doibase [0]{http://dx.doi.org/}%
\providecommand \selectlanguage [0]{\@gobble}%
\providecommand \bibinfo  [0]{\@secondoftwo}%
\providecommand \bibfield  [0]{\@secondoftwo}%
\providecommand \translation [1]{[#1]}%
\providecommand \BibitemOpen [0]{}%
\providecommand \bibitemStop [0]{}%
\providecommand \bibitemNoStop [0]{.\EOS\space}%
\providecommand \EOS [0]{\spacefactor3000\relax}%
\providecommand \BibitemShut  [1]{\csname bibitem#1\endcsname}%
\let\auto@bib@innerbib\@empty
\bibitem [{\citenamefont {Bohr}(1928)}]{bohr_quantum_1928}%
  \BibitemOpen
  \bibfield  {author} {\bibinfo {author} {\bibfnamefont {N.}~\bibnamefont
  {Bohr}},\ }\href {\doibase 10.1038/121580a0} {\bibfield  {journal} {\bibinfo
  {journal} {Nature}\ }\textbf {\bibinfo {volume} {121}},\ \bibinfo {pages}
  {580} (\bibinfo {year} {1928})}\BibitemShut {NoStop}%
\bibitem [{\citenamefont {Wootters}\ and\ \citenamefont
  {Zurek}(1979)}]{wootters_complementarity_1979}%
  \BibitemOpen
  \bibfield  {author} {\bibinfo {author} {\bibfnamefont {W.~K.}\ \bibnamefont
  {Wootters}}\ and\ \bibinfo {author} {\bibfnamefont {W.~H.}\ \bibnamefont
  {Zurek}},\ }\href {\doibase 10.1103/PhysRevD.19.473} {\bibfield  {journal}
  {\bibinfo  {journal} {Phys. Rev. D}\ }\textbf {\bibinfo {volume} {19}},\
  \bibinfo {pages} {473} (\bibinfo {year} {1979})}\BibitemShut {NoStop}%
\bibitem [{\citenamefont {Bartell}(1980)}]{bartell_complementarity_1980}%
  \BibitemOpen
  \bibfield  {author} {\bibinfo {author} {\bibfnamefont {L.~S.}\ \bibnamefont
  {Bartell}},\ }\href {\doibase 10.1103/PhysRevD.21.1698} {\bibfield  {journal}
  {\bibinfo  {journal} {Phys. Rev. D}\ }\textbf {\bibinfo {volume} {21}},\
  \bibinfo {pages} {1698} (\bibinfo {year} {1980})}\BibitemShut {NoStop}%
\bibitem [{\citenamefont {Greenberger}\ and\ \citenamefont
  {Yasin}(1988)}]{greenberger_simultaneous_1988}%
  \BibitemOpen
  \bibfield  {author} {\bibinfo {author} {\bibfnamefont {D.~M.}\ \bibnamefont
  {Greenberger}}\ and\ \bibinfo {author} {\bibfnamefont {A.}~\bibnamefont
  {Yasin}},\ }\href {\doibase 10.1016/0375-9601(88)90114-4} {\bibfield
  {journal} {\bibinfo  {journal} {Physics Letters A}\ }\textbf {\bibinfo
  {volume} {128}},\ \bibinfo {pages} {391} (\bibinfo {year}
  {1988})}\BibitemShut {NoStop}%
\bibitem [{\citenamefont {Englert}(1996)}]{englert_fringe_1996}%
  \BibitemOpen
  \bibfield  {author} {\bibinfo {author} {\bibfnamefont {B.-G.}\ \bibnamefont
  {Englert}},\ }\href {\doibase 10.1103/PhysRevLett.77.2154} {\bibfield
  {journal} {\bibinfo  {journal} {Phys. Rev. Lett.}\ }\textbf {\bibinfo
  {volume} {77}},\ \bibinfo {pages} {2154} (\bibinfo {year}
  {1996})}\BibitemShut {NoStop}%
\bibitem [{\citenamefont {Jaeger}\ \emph {et~al.}(1993)\citenamefont {Jaeger},
  \citenamefont {Horne},\ and\ \citenamefont
  {Shimony}}]{jaeger_complementarity_1993}%
  \BibitemOpen
  \bibfield  {author} {\bibinfo {author} {\bibfnamefont {G.}~\bibnamefont
  {Jaeger}}, \bibinfo {author} {\bibfnamefont {M.~A.}\ \bibnamefont {Horne}}, \
  and\ \bibinfo {author} {\bibfnamefont {A.}~\bibnamefont {Shimony}},\ }\href
  {\doibase 10.1103/PhysRevA.48.1023} {\bibfield  {journal} {\bibinfo
  {journal} {Phys. Rev. A}\ }\textbf {\bibinfo {volume} {48}},\ \bibinfo
  {pages} {1023} (\bibinfo {year} {1993})}\BibitemShut {NoStop}%
\bibitem [{\citenamefont {Jaeger}\ \emph {et~al.}(1995)\citenamefont {Jaeger},
  \citenamefont {Shimony},\ and\ \citenamefont {Vaidman}}]{jaeger_two_1995}%
  \BibitemOpen
  \bibfield  {author} {\bibinfo {author} {\bibfnamefont {G.}~\bibnamefont
  {Jaeger}}, \bibinfo {author} {\bibfnamefont {A.}~\bibnamefont {Shimony}}, \
  and\ \bibinfo {author} {\bibfnamefont {L.}~\bibnamefont {Vaidman}},\ }\href
  {\doibase 10.1103/PhysRevA.51.54} {\bibfield  {journal} {\bibinfo  {journal}
  {Phys. Rev. A}\ }\textbf {\bibinfo {volume} {51}},\ \bibinfo {pages} {54}
  (\bibinfo {year} {1995})}\BibitemShut {NoStop}%
\bibitem [{\citenamefont {Englert}\ and\ \citenamefont
  {Bergou}(2000)}]{englert_quantitative_2000}%
  \BibitemOpen
  \bibfield  {author} {\bibinfo {author} {\bibfnamefont {B.-G.}\ \bibnamefont
  {Englert}}\ and\ \bibinfo {author} {\bibfnamefont {J.~A.}\ \bibnamefont
  {Bergou}},\ }\href {\doibase 10.1016/S0030-4018(99)00718-X} {\bibfield
  {journal} {\bibinfo  {journal} {Optics Communications}\ }\textbf {\bibinfo
  {volume} {179}},\ \bibinfo {pages} {337} (\bibinfo {year}
  {2000})}\BibitemShut {NoStop}%
\bibitem [{\citenamefont {Martinez-Linares}\ and\ \citenamefont
  {Harmin}(2004)}]{martinez-linares_quality_2004}%
  \BibitemOpen
  \bibfield  {author} {\bibinfo {author} {\bibfnamefont {J.}~\bibnamefont
  {Martinez-Linares}}\ and\ \bibinfo {author} {\bibfnamefont {D.~A.}\
  \bibnamefont {Harmin}},\ }\href {\doibase 10.1103/PhysRevA.69.062109}
  {\bibfield  {journal} {\bibinfo  {journal} {Phys. Rev. A}\ }\textbf {\bibinfo
  {volume} {69}},\ \bibinfo {pages} {062109} (\bibinfo {year}
  {2004})}\BibitemShut {NoStop}%
\bibitem [{\citenamefont {Luis}(2004)}]{luis_operational_2004}%
  \BibitemOpen
  \bibfield  {author} {\bibinfo {author} {\bibfnamefont {A.}~\bibnamefont
  {Luis}},\ }\href {\doibase 10.1103/PhysRevA.70.062107} {\bibfield  {journal}
  {\bibinfo  {journal} {Phys. Rev. A}\ }\textbf {\bibinfo {volume} {70}},\
  \bibinfo {pages} {062107} (\bibinfo {year} {2004})}\BibitemShut {NoStop}%
\bibitem [{\citenamefont {Englert}\ \emph {et~al.}(2008)\citenamefont
  {Englert}, \citenamefont {Kaszlikowski}, \citenamefont {Kwek},\ and\
  \citenamefont {Chee}}]{englert_wave-particle_2008}%
  \BibitemOpen
  \bibfield  {author} {\bibinfo {author} {\bibfnamefont {B.-G.}\ \bibnamefont
  {Englert}}, \bibinfo {author} {\bibfnamefont {D.}~\bibnamefont
  {Kaszlikowski}}, \bibinfo {author} {\bibfnamefont {L.~C.}\ \bibnamefont
  {Kwek}}, \ and\ \bibinfo {author} {\bibfnamefont {W.~H.}\ \bibnamefont
  {Chee}},\ }\href {\doibase 10.1142/S0219749908003220} {\bibfield  {journal}
  {\bibinfo  {journal} {Int. J. Quantum Inform.}\ }\textbf {\bibinfo {volume}
  {06}},\ \bibinfo {pages} {129} (\bibinfo {year} {2008})}\BibitemShut
  {NoStop}%
\bibitem [{\citenamefont {Erez}\ \emph {et~al.}(2009)\citenamefont {Erez},
  \citenamefont {Jacobs},\ and\ \citenamefont
  {Kurizki}}]{erez_operational_2009}%
  \BibitemOpen
  \bibfield  {author} {\bibinfo {author} {\bibfnamefont {N.}~\bibnamefont
  {Erez}}, \bibinfo {author} {\bibfnamefont {D.}~\bibnamefont {Jacobs}}, \ and\
  \bibinfo {author} {\bibfnamefont {G.}~\bibnamefont {Kurizki}},\ }\href
  {\doibase 10.1088/0953-4075/42/11/114006} {\bibfield  {journal} {\bibinfo
  {journal} {J. Phys. B: At. Mol. Opt. Phys.}\ }\textbf {\bibinfo {volume}
  {42}},\ \bibinfo {pages} {114006} (\bibinfo {year} {2009})}\BibitemShut
  {NoStop}%
\bibitem [{\citenamefont {Liu}\ \emph {et~al.}(2009)\citenamefont {Liu},
  \citenamefont {Li}, \citenamefont {Yu},\ and\ \citenamefont
  {Chen}}]{liu_duality_2009}%
  \BibitemOpen
  \bibfield  {author} {\bibinfo {author} {\bibfnamefont {N.-L.}\ \bibnamefont
  {Liu}}, \bibinfo {author} {\bibfnamefont {L.}~\bibnamefont {Li}}, \bibinfo
  {author} {\bibfnamefont {S.}~\bibnamefont {Yu}}, \ and\ \bibinfo {author}
  {\bibfnamefont {Z.-B.}\ \bibnamefont {Chen}},\ }\href {\doibase
  10.1103/PhysRevA.79.052108} {\bibfield  {journal} {\bibinfo  {journal} {Phys.
  Rev. A}\ }\textbf {\bibinfo {volume} {79}},\ \bibinfo {pages} {052108}
  (\bibinfo {year} {2009})}\BibitemShut {NoStop}%
\bibitem [{\citenamefont {Li}\ \emph {et~al.}(2012)\citenamefont {Li},
  \citenamefont {Liu},\ and\ \citenamefont {Yu}}]{li_duality_2012}%
  \BibitemOpen
  \bibfield  {author} {\bibinfo {author} {\bibfnamefont {L.}~\bibnamefont
  {Li}}, \bibinfo {author} {\bibfnamefont {N.-L.}\ \bibnamefont {Liu}}, \ and\
  \bibinfo {author} {\bibfnamefont {S.}~\bibnamefont {Yu}},\ }\href {\doibase
  10.1103/PhysRevA.85.054101} {\bibfield  {journal} {\bibinfo  {journal} {Phys.
  Rev. A}\ }\textbf {\bibinfo {volume} {85}},\ \bibinfo {pages} {054101}
  (\bibinfo {year} {2012})}\BibitemShut {NoStop}%
\bibitem [{\citenamefont {Coles}\ \emph {et~al.}(2014)\citenamefont {Coles},
  \citenamefont {Kaniewski},\ and\ \citenamefont
  {Wehner}}]{coles_equivalence_2014}%
  \BibitemOpen
  \bibfield  {author} {\bibinfo {author} {\bibfnamefont {P.~J.}\ \bibnamefont
  {Coles}}, \bibinfo {author} {\bibfnamefont {J.}~\bibnamefont {Kaniewski}}, \
  and\ \bibinfo {author} {\bibfnamefont {S.}~\bibnamefont {Wehner}},\ }\href
  {\doibase 10.1038/ncomms6814} {\bibfield  {journal} {\bibinfo  {journal} {Nat
  Commun}\ }\textbf {\bibinfo {volume} {5}},\ \bibinfo {pages} {5814} (\bibinfo
  {year} {2014})}\BibitemShut {NoStop}%
\bibitem [{\citenamefont {Jia}\ \emph {et~al.}(2014)\citenamefont {Jia},
  \citenamefont {Huang}, \citenamefont {Zhang},\ and\ \citenamefont
  {Zhu}}]{jia_influence_2014}%
  \BibitemOpen
  \bibfield  {author} {\bibinfo {author} {\bibfnamefont {A.-A.}\ \bibnamefont
  {Jia}}, \bibinfo {author} {\bibfnamefont {J.-H.}\ \bibnamefont {Huang}},
  \bibinfo {author} {\bibfnamefont {T.-C.}\ \bibnamefont {Zhang}}, \ and\
  \bibinfo {author} {\bibfnamefont {S.-Y.}\ \bibnamefont {Zhu}},\ }\href
  {\doibase 10.1103/PhysRevA.89.042103} {\bibfield  {journal} {\bibinfo
  {journal} {Phys. Rev. A}\ }\textbf {\bibinfo {volume} {89}},\ \bibinfo
  {pages} {042103} (\bibinfo {year} {2014})}\BibitemShut {NoStop}%
\bibitem [{\citenamefont {Bera}\ \emph {et~al.}(2015)\citenamefont {Bera},
  \citenamefont {Qureshi}, \citenamefont {Siddiqui},\ and\ \citenamefont
  {Pati}}]{bera_duality_2015}%
  \BibitemOpen
  \bibfield  {author} {\bibinfo {author} {\bibfnamefont {M.~N.}\ \bibnamefont
  {Bera}}, \bibinfo {author} {\bibfnamefont {T.}~\bibnamefont {Qureshi}},
  \bibinfo {author} {\bibfnamefont {M.~A.}\ \bibnamefont {Siddiqui}}, \ and\
  \bibinfo {author} {\bibfnamefont {A.~K.}\ \bibnamefont {Pati}},\ }\href
  {\doibase 10.1103/PhysRevA.92.012118} {\bibfield  {journal} {\bibinfo
  {journal} {Phys. Rev. A}\ }\textbf {\bibinfo {volume} {92}},\ \bibinfo
  {pages} {012118} (\bibinfo {year} {2015})}\BibitemShut {NoStop}%
\bibitem [{\citenamefont {Rauch}\ and\ \citenamefont
  {Summhammer}(1984)}]{rauch_static_1984}%
  \BibitemOpen
  \bibfield  {author} {\bibinfo {author} {\bibfnamefont {H.}~\bibnamefont
  {Rauch}}\ and\ \bibinfo {author} {\bibfnamefont {J.}~\bibnamefont
  {Summhammer}},\ }\href {\doibase 10.1016/0375-9601(84)90586-3} {\bibfield
  {journal} {\bibinfo  {journal} {Physics Letters A}\ }\textbf {\bibinfo
  {volume} {104}},\ \bibinfo {pages} {44} (\bibinfo {year} {1984})}\BibitemShut
  {NoStop}%
\bibitem [{\citenamefont {Summhammer}\ \emph {et~al.}(1987)\citenamefont
  {Summhammer}, \citenamefont {Rauch},\ and\ \citenamefont
  {Tuppinger}}]{summhammer_stochastic_1987}%
  \BibitemOpen
  \bibfield  {author} {\bibinfo {author} {\bibfnamefont {J.}~\bibnamefont
  {Summhammer}}, \bibinfo {author} {\bibfnamefont {H.}~\bibnamefont {Rauch}}, \
  and\ \bibinfo {author} {\bibfnamefont {D.}~\bibnamefont {Tuppinger}},\ }\href
  {\doibase 10.1103/PhysRevA.36.4447} {\bibfield  {journal} {\bibinfo
  {journal} {Phys. Rev. A}\ }\textbf {\bibinfo {volume} {36}},\ \bibinfo
  {pages} {4447} (\bibinfo {year} {1987})}\BibitemShut {NoStop}%
\bibitem [{\citenamefont {Liu}\ \emph {et~al.}(2012)\citenamefont {Liu},
  \citenamefont {Huang}, \citenamefont {Gao}, \citenamefont {Zubairy},\ and\
  \citenamefont {Zhu}}]{liu_relation_2012}%
  \BibitemOpen
  \bibfield  {author} {\bibinfo {author} {\bibfnamefont {H.-Y.}\ \bibnamefont
  {Liu}}, \bibinfo {author} {\bibfnamefont {J.-H.}\ \bibnamefont {Huang}},
  \bibinfo {author} {\bibfnamefont {J.-R.}\ \bibnamefont {Gao}}, \bibinfo
  {author} {\bibfnamefont {M.~S.}\ \bibnamefont {Zubairy}}, \ and\ \bibinfo
  {author} {\bibfnamefont {S.-Y.}\ \bibnamefont {Zhu}},\ }\href {\doibase
  10.1103/PhysRevA.85.022106} {\bibfield  {journal} {\bibinfo  {journal} {Phys.
  Rev. A}\ }\textbf {\bibinfo {volume} {85}},\ \bibinfo {pages} {022106}
  (\bibinfo {year} {2012})}\BibitemShut {NoStop}%
\bibitem [{\citenamefont {Tang}\ \emph {et~al.}(2013)\citenamefont {Tang},
  \citenamefont {Li}, \citenamefont {Li},\ and\ \citenamefont
  {Guo}}]{tang_revisiting_2013}%
  \BibitemOpen
  \bibfield  {author} {\bibinfo {author} {\bibfnamefont {J.-S.}\ \bibnamefont
  {Tang}}, \bibinfo {author} {\bibfnamefont {Y.-L.}\ \bibnamefont {Li}},
  \bibinfo {author} {\bibfnamefont {C.-F.}\ \bibnamefont {Li}}, \ and\ \bibinfo
  {author} {\bibfnamefont {G.-C.}\ \bibnamefont {Guo}},\ }\href {\doibase
  10.1103/PhysRevA.88.014103} {\bibfield  {journal} {\bibinfo  {journal} {Phys.
  Rev. A}\ }\textbf {\bibinfo {volume} {88}},\ \bibinfo {pages} {014103}
  (\bibinfo {year} {2013})}\BibitemShut {NoStop}%
\bibitem [{\citenamefont {Yan}\ \emph {et~al.}(2015)\citenamefont {Yan},
  \citenamefont {Liao}, \citenamefont {Deng}, \citenamefont {He}, \citenamefont
  {Xue}, \citenamefont {Zhang},\ and\ \citenamefont
  {Zhu}}]{yan_experimental_2015}%
  \BibitemOpen
  \bibfield  {author} {\bibinfo {author} {\bibfnamefont {H.}~\bibnamefont
  {Yan}}, \bibinfo {author} {\bibfnamefont {K.}~\bibnamefont {Liao}}, \bibinfo
  {author} {\bibfnamefont {Z.}~\bibnamefont {Deng}}, \bibinfo {author}
  {\bibfnamefont {J.}~\bibnamefont {He}}, \bibinfo {author} {\bibfnamefont
  {Z.-Y.}\ \bibnamefont {Xue}}, \bibinfo {author} {\bibfnamefont {Z.-M.}\
  \bibnamefont {Zhang}}, \ and\ \bibinfo {author} {\bibfnamefont {S.-L.}\
  \bibnamefont {Zhu}},\ }\href {\doibase 10.1103/PhysRevA.91.042132} {\bibfield
   {journal} {\bibinfo  {journal} {Phys. Rev. A}\ }\textbf {\bibinfo {volume}
  {91}},\ \bibinfo {pages} {042132} (\bibinfo {year} {2015})}\BibitemShut
  {NoStop}%
\bibitem [{\citenamefont {Heuer}\ \emph {et~al.}(2015)\citenamefont {Heuer},
  \citenamefont {Pieplow},\ and\ \citenamefont
  {Menzel}}]{heuer_phase-selective_2015}%
  \BibitemOpen
  \bibfield  {author} {\bibinfo {author} {\bibfnamefont {A.}~\bibnamefont
  {Heuer}}, \bibinfo {author} {\bibfnamefont {G.}~\bibnamefont {Pieplow}}, \
  and\ \bibinfo {author} {\bibfnamefont {R.}~\bibnamefont {Menzel}},\ }\href
  {\doibase 10.1103/PhysRevA.92.013803} {\bibfield  {journal} {\bibinfo
  {journal} {Phys. Rev. A}\ }\textbf {\bibinfo {volume} {92}},\ \bibinfo
  {pages} {013803} (\bibinfo {year} {2015})}\BibitemShut {NoStop}%
\bibitem [{\citenamefont {Jacques}\ \emph {et~al.}(2007)\citenamefont
  {Jacques}, \citenamefont {Wu}, \citenamefont {Grosshans}, \citenamefont
  {Treussart}, \citenamefont {Grangier}, \citenamefont {Aspect},\ and\
  \citenamefont {Roch}}]{jacques_experimental_2007}%
  \BibitemOpen
  \bibfield  {author} {\bibinfo {author} {\bibfnamefont {V.}~\bibnamefont
  {Jacques}}, \bibinfo {author} {\bibfnamefont {E.}~\bibnamefont {Wu}},
  \bibinfo {author} {\bibfnamefont {F.}~\bibnamefont {Grosshans}}, \bibinfo
  {author} {\bibfnamefont {F.}~\bibnamefont {Treussart}}, \bibinfo {author}
  {\bibfnamefont {P.}~\bibnamefont {Grangier}}, \bibinfo {author}
  {\bibfnamefont {A.}~\bibnamefont {Aspect}}, \ and\ \bibinfo {author}
  {\bibfnamefont {J.-F.}\ \bibnamefont {Roch}},\ }\href {\doibase
  10.1126/science.1136303} {\bibfield  {journal} {\bibinfo  {journal}
  {Science}\ }\textbf {\bibinfo {volume} {315}},\ \bibinfo {pages} {966}
  (\bibinfo {year} {2007})}\BibitemShut {NoStop}%
\bibitem [{\citenamefont {Jacques}\ \emph {et~al.}(2008)\citenamefont
  {Jacques}, \citenamefont {Wu}, \citenamefont {Grosshans}, \citenamefont
  {Treussart}, \citenamefont {Grangier}, \citenamefont {Aspect},\ and\
  \citenamefont {Roch}}]{jacques_delayed-choice_2008}%
  \BibitemOpen
  \bibfield  {author} {\bibinfo {author} {\bibfnamefont {V.}~\bibnamefont
  {Jacques}}, \bibinfo {author} {\bibfnamefont {E.}~\bibnamefont {Wu}},
  \bibinfo {author} {\bibfnamefont {F.}~\bibnamefont {Grosshans}}, \bibinfo
  {author} {\bibfnamefont {F.}~\bibnamefont {Treussart}}, \bibinfo {author}
  {\bibfnamefont {P.}~\bibnamefont {Grangier}}, \bibinfo {author}
  {\bibfnamefont {A.}~\bibnamefont {Aspect}}, \ and\ \bibinfo {author}
  {\bibfnamefont {J.-F.}\ \bibnamefont {Roch}},\ }\href {\doibase
  10.1103/PhysRevLett.100.220402} {\bibfield  {journal} {\bibinfo  {journal}
  {Phys. Rev. Lett.}\ }\textbf {\bibinfo {volume} {100}},\ \bibinfo {pages}
  {220402} (\bibinfo {year} {2008})}\BibitemShut {NoStop}%
\bibitem [{\citenamefont {Manning}\ \emph {et~al.}(2015)\citenamefont
  {Manning}, \citenamefont {Khakimov}, \citenamefont {Dall},\ and\
  \citenamefont {Truscott}}]{manning_wheelers_2015}%
  \BibitemOpen
  \bibfield  {author} {\bibinfo {author} {\bibfnamefont {A.~G.}\ \bibnamefont
  {Manning}}, \bibinfo {author} {\bibfnamefont {R.~I.}\ \bibnamefont
  {Khakimov}}, \bibinfo {author} {\bibfnamefont {R.~G.}\ \bibnamefont {Dall}},
  \ and\ \bibinfo {author} {\bibfnamefont {A.~G.}\ \bibnamefont {Truscott}},\
  }\href {\doibase 10.1038/nphys3343} {\bibfield  {journal} {\bibinfo
  {journal} {Nat Phys}\ }\textbf {\bibinfo {volume} {11}},\ \bibinfo {pages}
  {539} (\bibinfo {year} {2015})}\BibitemShut {NoStop}%
\bibitem [{\citenamefont {Arndt}\ \emph {et~al.}(1999)\citenamefont {Arndt},
  \citenamefont {Nairz}, \citenamefont {Vos-Andreae}, \citenamefont {Keller},
  \citenamefont {van~der Zouw},\ and\ \citenamefont
  {Zeilinger}}]{arndt_waveparticle_1999}%
  \BibitemOpen
  \bibfield  {author} {\bibinfo {author} {\bibfnamefont {M.}~\bibnamefont
  {Arndt}}, \bibinfo {author} {\bibfnamefont {O.}~\bibnamefont {Nairz}},
  \bibinfo {author} {\bibfnamefont {J.}~\bibnamefont {Vos-Andreae}}, \bibinfo
  {author} {\bibfnamefont {C.}~\bibnamefont {Keller}}, \bibinfo {author}
  {\bibfnamefont {G.}~\bibnamefont {van~der Zouw}}, \ and\ \bibinfo {author}
  {\bibfnamefont {A.}~\bibnamefont {Zeilinger}},\ }\href {\doibase
  10.1038/44348} {\bibfield  {journal} {\bibinfo  {journal} {Nature}\ }\textbf
  {\bibinfo {volume} {401}},\ \bibinfo {pages} {680} (\bibinfo {year}
  {1999})}\BibitemShut {NoStop}%
\bibitem [{\citenamefont {Couder}\ \emph {et~al.}(2005)\citenamefont {Couder},
  \citenamefont {Protière}, \citenamefont {Fort},\ and\ \citenamefont
  {Boudaoud}}]{couder_dynamical_2005}%
  \BibitemOpen
  \bibfield  {author} {\bibinfo {author} {\bibfnamefont {Y.}~\bibnamefont
  {Couder}}, \bibinfo {author} {\bibfnamefont {S.}~\bibnamefont {Protière}},
  \bibinfo {author} {\bibfnamefont {E.}~\bibnamefont {Fort}}, \ and\ \bibinfo
  {author} {\bibfnamefont {A.}~\bibnamefont {Boudaoud}},\ }\href {\doibase
  10.1038/437208a} {\bibfield  {journal} {\bibinfo  {journal} {Nature}\
  }\textbf {\bibinfo {volume} {437}},\ \bibinfo {pages} {208} (\bibinfo {year}
  {2005})}\BibitemShut {NoStop}%
\bibitem [{\citenamefont {Couder}\ and\ \citenamefont
  {Fort}(2006)}]{couder_single-particle_2006}%
  \BibitemOpen
  \bibfield  {author} {\bibinfo {author} {\bibfnamefont {Y.}~\bibnamefont
  {Couder}}\ and\ \bibinfo {author} {\bibfnamefont {E.}~\bibnamefont {Fort}},\
  }\href {\doibase 10.1103/PhysRevLett.97.154101} {\bibfield  {journal}
  {\bibinfo  {journal} {Phys. Rev. Lett.}\ }\textbf {\bibinfo {volume} {97}},\
  \bibinfo {pages} {154101} (\bibinfo {year} {2006})}\BibitemShut {NoStop}%
\bibitem [{\citenamefont {Bush}(2010)}]{bush_quantum_2010}%
  \BibitemOpen
  \bibfield  {author} {\bibinfo {author} {\bibfnamefont {J.~W.~M.}\
  \bibnamefont {Bush}},\ }\href {\doibase 10.1073/pnas.1012399107} {\bibfield
  {journal} {\bibinfo  {journal} {PNAS}\ }\textbf {\bibinfo {volume} {107}},\
  \bibinfo {pages} {17455} (\bibinfo {year} {2010})}\BibitemShut {NoStop}%
\end{thebibliography}%

\end{document}